\documentclass[aps, pra, a4paper,showpacs,twocolumn, 10pt]{revtex4-1}
\usepackage{bbm, amsmath, amssymb, amsthm, bm,textcomp, nicefrac, mathtools,geometry}
\usepackage{amsthm,amssymb,amsmath,graphicx,epsfig,color,verbatim,enumerate,amsthm}

\newcommand{\ket}[1]{\left|#1\right\rangle}

\newcommand\defn[1]{\textsl{#1}}

\newcommand{\C}{\ensuremath{\mathbbm C}}
\newcommand{\be}{\begin{equation}}
\newcommand{\ee}{\end{equation}}

\newcommand{\bea}{\begin{eqnarray}}
\newcommand{\eea}{\end{eqnarray}}

\newcommand{\one}{\mbox{$1 \hspace{-1.0mm}  {\bf l}$}}

\makeatletter
\newtheorem*{rep@theorem}{\rep@title}
\newcommand{\newreptheorem}[2]{%
\newenvironment{rep#1}[1]{%
 \def\rep@title{#2 \ref{##1}}%
 \begin{rep@theorem}}%
 {\end{rep@theorem}}}
\makeatother

\newreptheorem{theorem}{Theorem}

\begin{document}
\title{Deterministic super-replication of one-parameter unitary transformations}
\author{W. D\"ur, P. Sekatski and M. Skotiniotis}

\affiliation{Institut f\"ur Theoretische Physik, Universit\"at
  Innsbruck, Technikerstr. 25, A-6020 Innsbruck,  Austria}
\date{\today}

\begin{abstract}
We show that one can deterministically generate out of $N$ copies of an unknown unitary operation up to $N^2$ almost perfect copies. The result holds for all operations generated by a Hamiltonian with an unknown interaction strength. This generalizes a similar result in the context of phase covariant cloning where, however, super-replication comes at the price of an exponentially reduced probability of success. We also show that multiple copies of unitary operations can be emulated by operations acting on a much smaller space, e.g.,~a  magnetic field acting on a single $n$-level system allows one to emulate the action of the field on $n^2$ qubits.
\end{abstract}
\pacs{03.67.-a, 03.65.Ud, 03.67.Hk, 03.65.Ta}
\maketitle

{\em Introduction.---}
Quantum information can not be cloned. This simple statement, first manifested in~\cite{Wo:82}, has far reaching
consequences particularly in the context of quantum cryptography where the no-cloning principle ensures
security~\cite{Qcrypto}. A violation of the no-cloning principle would allow for super-luminal communication or the violation of Heisenberg's uncertainty principle, illustrating its fundamental character.

However, imperfect replication of quantum information is possible and various works have derived optimal cloning
devices under different circumstances~\cite{Buzek:96,*Gisin:97,*Bruss:98, *Werner:98}
(see also~\cite{Scarani:05,*Fan:14}). Given $N$ copies of a system in some pure state $\ket{\psi}$, one can
deterministically produce $M>N$ copies with a non-unit fidelity that depends on $N$ and $M$.
Moreover, it was shown in~\cite{Ch:13} that for \defn{phase covariant states}, i.e., states generated by a Hamiltonian with unknown interaction strength, up to $N^2$ almost perfect copies can be generated using a probabilistic replication processes. 
This \defn{super-replication} of states comes at the price of a success probability that drops exponentially with $N$.

In this letter we show that a similar super-replication can be achieved for the cloning of unitary operations. Here the
goal is to produce out of $N$ copies of an unknown unitary operation (given in the form of a black box that can be
applied to arbitrary states) $M \geq N$ copies. This is in general a harder task than cloning of
states, as one needs to replicate the action of the operation on all possible input states~\cite{Ch:08}. Nevertheless, we
find that {\em deterministic} super-replication of unitary transformations of the form $U=e^{i\vartheta H}$, where $H$ is
the Hamiltonian generating the unitary evolution and $\vartheta$ is an unknown interaction strength, is possible in
contrast to super-replication of phase covariant states. We demonstrate this result by providing an explicit protocol that
makes use of likely sequences as in Schumacher's compression theorem \cite{Sch:95}.

We also consider the emulation of multiple copies of unitary operations by operations acting on a smaller space.
Specifically, we find that a single unitary operation with a given interaction strength, $\vartheta$, acting on an
$n$-dimensional system is sufficient to emulate $n^2$ copies of an operation with the same $\vartheta$ acting on a
two-dimensional system. In addition, we show that if one can also interject the evolution generated by the Hamiltonian
with additional control operations, then one can generate an arbitrary number, $M$, of perfect copies from single
instance of the unitary operation at the cost of a  $\sqrt M$ reduction in interaction strength.

{\em Background.---}
We start by specifying the set-up. We consider a class of $d$-dimensional unitary operations
$U(\vartheta)=\exp({-i \vartheta H})$ generated by a Hamiltonian, $H$, and parametrized by $\vartheta$. The
operations are provided in the form of an unknown black box, where
$H=\textstyle{\sum_{k=0}^{d-1} \lambda_k |\varphi_k\rangle\langle \varphi_k|}$ is known but $\vartheta$ is not.
For instance, this may correspond to a situation where $\vartheta$ specifies the unknown interaction strength and the time for
applying $H$ is fixed. For simplicity we will consider $d=2$ in the following where $H = |1\rangle\langle 1|$
and $\vartheta \in [0,2\pi)$, i.e., $U(\vartheta)$ is equivalent, up to an irrelevant global phase factor, to a rotation around the
$z$-axis by an angle $\vartheta$. Generalization of the results to arbitrary $d$ and arbitrary Hamiltonians are
straightforward.

\begin{figure}[ht]
\begin{picture}(210,120)
\put(-22,-60){\includegraphics[width=8.5cm]{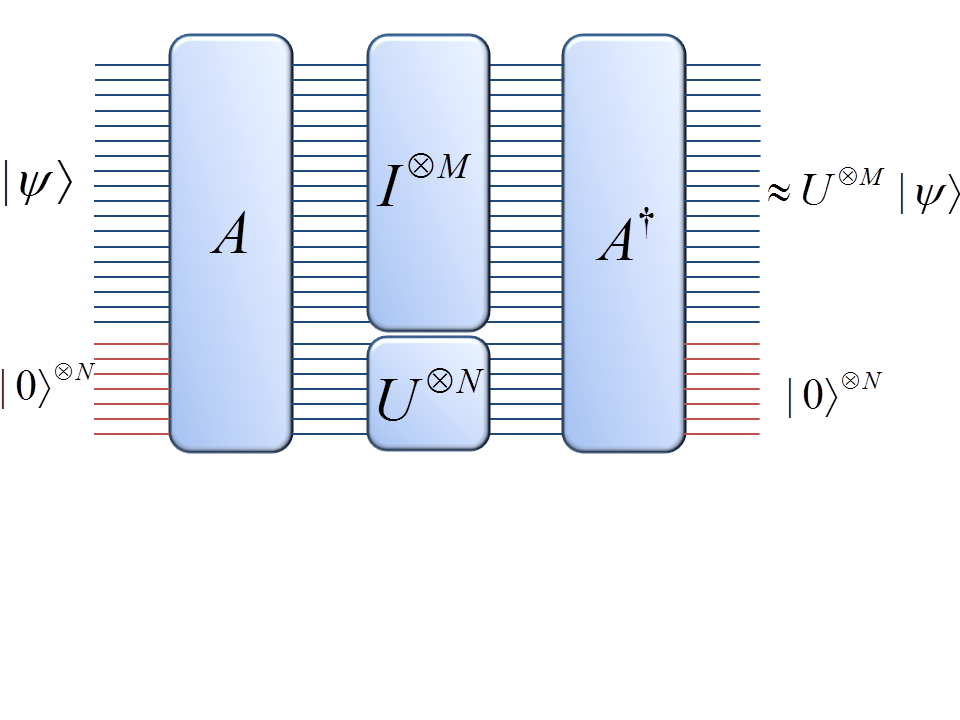}}
\end{picture}
\caption{Illustration of the overall procedure to obtain $M$ approximate copies from $N$ applications of an (unknown) unitary operation $U(\vartheta)$. By applying the unitary basis change $A^\dagger$, $A$, before and after $U(\vartheta)^{\otimes N}$ we can obtain the operation $\tilde{V}(\vartheta)$ of Eq.~\eqref{tildeV} which is a good approximation of $U(\vartheta)^{\otimes M}$, i.e., $A^\dagger (\one^{\otimes M} \otimes U^{\otimes N}) A\left(|\psi\rangle\otimes|0\rangle^{\otimes N}\right)\approx U^{\otimes M}|\psi\rangle \otimes |0\rangle^{\otimes N}$. Notice that $N$ auxiliary systems are used that are not affected by the transformation.}
\label{Fig_Setup}
\end{figure}

The goal is to generate $M$ approximate copies of $U(\vartheta)$, i.e., $\tilde{V}(\vartheta)\approx U(\vartheta)^{\otimes M}$, given only $N$ copies, where
$M\geq N$.  To achieve this task we make use of a suitable number of auxiliary qubits and appropriate unitary operation $A$---to be applied before and after the application of $U(\vartheta)^{\otimes N}$---that yield an approximation of
$U(\vartheta)^{\otimes M}$ on an arbitrary input state, see Fig. \ref{Fig_Setup}.

We quantify the performance of $\tilde{V}(\vartheta)$ resulting from our protocol by the
\defn{global Jamio\l{}kowski Fidelity} (process fidelity), $F_E$, averaged over all possible input operations
$U(\vartheta)$~\cite{Ch:08,Gilchrist:05,*Dur:05}. For an $n$-dimensional unitary operation, $X$, the process fidelity of a completely
positive map $\cal E$ is defined as
\bea
F_E({\cal E},X) &=& \langle \psi_X| \rho_E |\psi_X\rangle
\eea
where $|\psi_X\rangle$, $\rho_E$ are the Choi-Jamio\l{}kowski states associated to $X$, ${\cal E}$ respectively via the
Choi-Jamio\l{}kowski isomorphism~\cite{Jamiolkowski:72,*Choi:72}. The latter associates to the operations $X$, ${\cal E}$ the states
$|\psi_X\rangle = \one \otimes X|\Phi\rangle$ and $\rho_E = \hat\one \otimes {\cal E}\left( |\Phi\rangle\langle \Phi|\right)$
respectively, where  $|\Phi\rangle=1/\sqrt{n}\sum_{j=1}^n |j\rangle\otimes |j\rangle$ is a maximally entangled $n$-level
state.

The process fidelity is closely related to the average fidelity,
$\bar F({\cal E},X)= \int {\text d} \psi \langle \psi|U^\dagger {\cal E}(|\psi\rangle\langle \psi|) U|\psi\rangle$, where the
average is taken over all input states $|\psi\rangle$. It is known that
$\bar F({\cal E},X) = (F_E({\cal E},X) n + 1)/(n+1)$~\cite{HHH:99,*Nielsen:02}, meaning that a sufficiently large process fidelity
ensures that the
map ${\cal E}$ provides a good approximation, on average, for all input states. Throughout this article we consider only
unitary operations, where the process fidelity reduces to the overlap of the corresponding pure Jamio\l{}kowski states.

{\em Faithful approximation of $U(\vartheta)^{\otimes M}$.---}
Consider $M$ copies of an operation
\be
\label{Utheta}U(\vartheta)=e^{-i \vartheta |1\rangle \langle 1|} = |0\rangle\langle 0| + e^{-i \vartheta} |1 \rangle\langle 1|.
\ee
We have that
\be
\label{UM}
U(\vartheta)^{\otimes M}=\sum_{\bm k} e^{-i |{\bm k}| \vartheta}|{\bm k}\rangle\langle {\bm k}|,
\ee
where we denote by $|{\bm k}\rangle \in (\C^2)^{\otimes M}$ the basis vectors of the $M$-qubit systems using binary
notation, i.e., $|{\bm 0}\rangle = |00 \ldots 0\rangle$, and $|{\bm k}|$ denotes the \defn{Hamming weight} of the vector $
{\bm k}$---the number of ones in binary notation. The corresponding Jamio\l{}kowski state,
$\one \otimes U(\vartheta)^{\otimes M} |\Phi\rangle$, with
$|\Phi\rangle = 2^{-M/2} \sum_{\bm k} |{\bm k}\rangle \otimes |{\bm k}\rangle$ is given by
\bea
|\psi_{U(\vartheta)^{\otimes M}}\rangle = 2^{-M/2} \sum_{\bm k}  e^{-i |{\bm k}| \vartheta} |{\bm k}\rangle \otimes |{\bm k}\rangle,
\eea
and all basis vectors with the same Hamming weight pick up the same phase factor.

Our goal is to approximate the action of $U(\vartheta)^{\otimes M}$. To this aim, consider an
operation $\tilde{V}(\vartheta)$ acting on $M$ qubits that only produces the appropriate phases for the majority of basis
vectors. The underlying distribution of the basis vectors in $U(\vartheta)^{\otimes M}$ is binomial, centered at $k=|{\bm k}|=M/2$, and in the limit of large $M$ approaches the
Gaussian distribution of the same mean and standard deviation $\sigma=\sqrt{M}/2$~\footnote{We note that, whereas the binomial distribution is strictly defined over the positive real line, in the limit of large $M$ it
can be shown that the Gaussian distribution over the negative real numbers only incurs an error in the approximation
that scales as $\mathcal{O}(M^{-1})$~\cite{SG:12}}. Hence, it suffices to reproduce
phases for $k \in (k_-,k_+)$ with $k_\pm =M/2 \pm \alpha M^\beta$ for some $\alpha>0$ and $1/2 < \beta <1$. The
operation
\be
\label{tildeV}
\tilde{V}(\vartheta) = \sum_{|{\bm k}| \in (k_-,k_+)} e^{-i |{\bm k}| \vartheta} |{\bm k}\rangle \langle {\bm k}| + \sum_{|{\bm k}| \not\in
(k_-,k_+)} e^{-i\gamma_{\bm k}} |{\bm k}\rangle \langle {\bm k}|,
\ee
with arbitrary $\gamma_{\bm k}$ approximates $U(\vartheta)^{\otimes M}$, where the process fidelity $F_E(\tilde{V}
(\vartheta), U(\vartheta)^{\otimes M}) = |\langle \psi_{\tilde{V}(\vartheta)}| \psi_{U(\vartheta)^{\otimes M}}\rangle |^2$ is
bounded from below by $\Phi(2 \alpha M^{\beta -1/2})= 1/\sqrt{2 \pi} \int_{- \alpha M^{\beta -1/2}}^{ \alpha M^{\beta -1/2}} e^{y^2/2} {\text d} y$
for any value of $\vartheta$. For our choice of $\alpha,\beta$, we have that $F_E \to 1$ for large $M$.
Notice that also for finite, moderate values of of $M$ one obtains a faithful approximation, which can be checked by directly evaluating the sum of binomial coefficients. Using Stirling's formula, one can approximate the binomial coefficients directly instead of invoking the
Gaussian approximation, and arrives at the same conclusion, i.e. for our choice of $\alpha,\,\beta$, $F_E \to 1$ in the limit of large $M$.

{\em Cloning protocol.---}
We now show how to obtain $\tilde V(\vartheta) \approx U(\vartheta)^{\otimes M}$ from $U(\vartheta)^{\otimes N}$
whenever $N=M^\beta, \forall \beta>1/2$.  As mentioned above it is sufficient to obtain the proper phases on all basis
states $|{\bm k}\rangle$ with $|{\bm k}| \in (k_-,k_+)$.  The latter set contains $2 \alpha M^\beta+1$ different phases,
with values $k_-+m\vartheta$ where $0 \leq m \leq 2 \alpha M^\beta$. Furthermore, we need only reproduce the
phases $m\vartheta$ as the resulting operation is equivalent up to an irrelevant  global phase factor $e^{-i k_- \vartheta}$.
As $U(\vartheta)^{\otimes N}$ contains $N+1$ distinct phases, $e^{i |{\bm k}|\vartheta}, \quad 0 \leq |{\bm k}| \leq N$
(see Eq.~\eqref{UM}), choosing $N = 2 \alpha M^\beta$ is sufficient to reproduce all the required phases of
$\tilde{V}(\vartheta)$ in the interval $(k_-,k_+)$ (see Eq.~\eqref{tildeV}).

\begin{figure}[ht]
\begin{center}
\begin{picture}(210,125)
\put(-22,-20){\includegraphics[width=8cm]{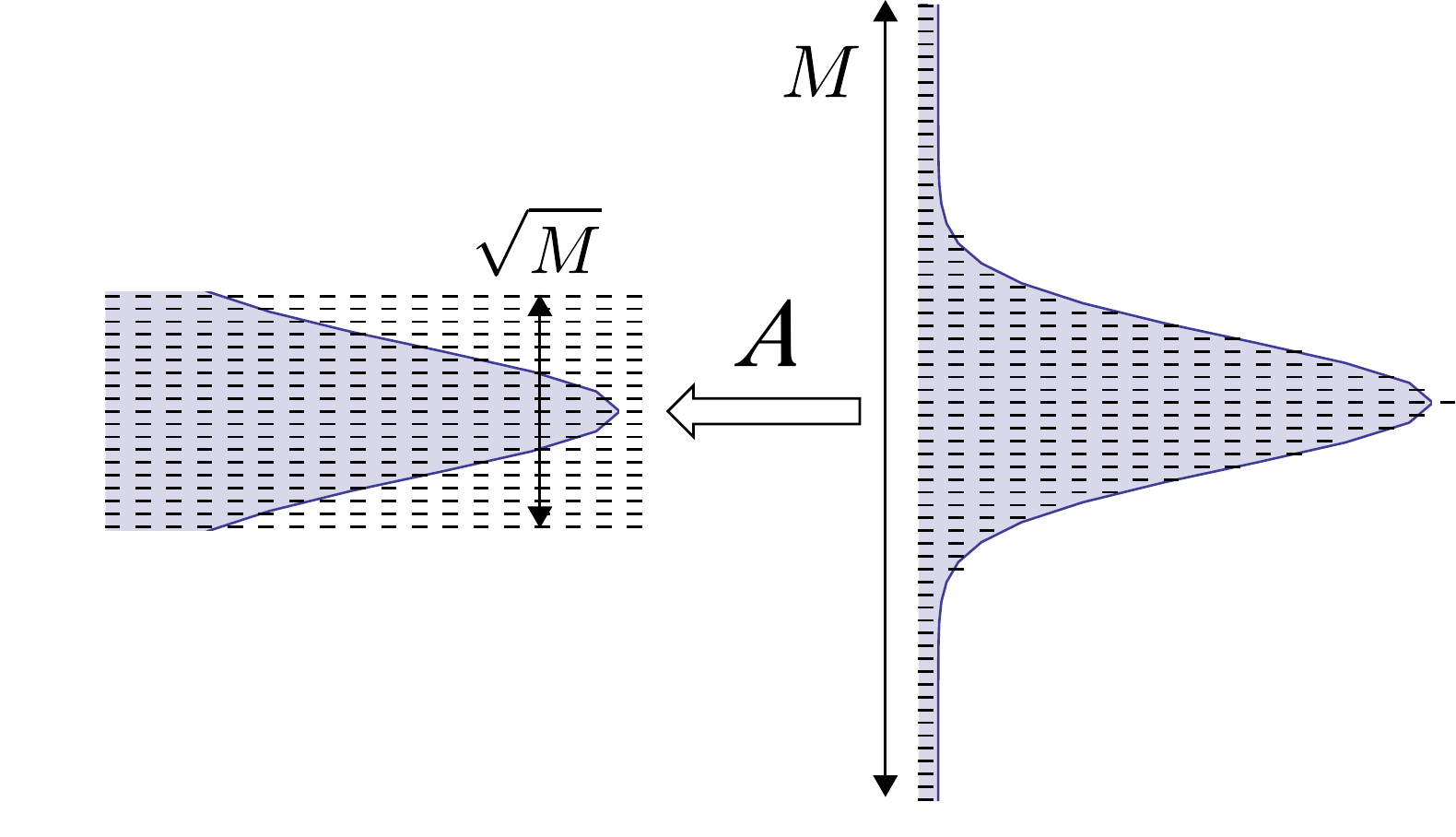}}
\end{picture}
\end{center}
\caption{Illustration of the unitary mapping $A$. Eigenstates $|{\bm k}\rangle$ of $U^{\otimes M}$ with corresponding degeneracies are depicted on the right. Eigenstates $|{\bm j}\rangle$ of $U(\vartheta)^{\otimes N}$ are depicted on the left, where the degeneracy are achieved by adding $M$ ancillay systems on which an identity operation acts. The relevant part of the spectrum of $U(\vartheta)^{\otimes M}$ is mapped to the appropriate eigenstates of $U^{\otimes N}\otimes I^{\otimes M}$.}
\label{Fig_Spectrum}
\end{figure}

To properly approximate $U(\vartheta)^{\otimes M}$ each phase $e^{-i |{\bm k}| \vartheta}$ has to be reproduced on all the $\binom{M}{|{\bm k}|}$ levels that lay in the multiplicity space for each $|{\bm k}|\in(k_-,k_+)$. To do so, we attach $M$ additional auxiliary systems
and consider the operation $\one^{\otimes M} \otimes U(\vartheta)^{\otimes N}$ (see Fig.~\ref{Fig_Spectrum}).  As the largest
multiplicity in $\tilde{V}(\vartheta)$ is $\binom{M}{M/2}$, $M$ auxiliary systems are sufficient as each eigenstate in
$\one^{\otimes M} \otimes U(\vartheta)^{\otimes N}$ is $2^M\binom{N}{|\bm k|}$-degenerate.

To obtain $\tilde{V}(\vartheta)$ from $\one^{\otimes M} \otimes U(\vartheta)^{\otimes N}$, all we need is to establish a
basis change that maps the eigenstates with the appropriate phases onto each other.  This is done as follows.
Consider the $M+N$ qubit state $\ket{\bm k}\otimes \ket{\bm 0}$ where $\ket{\bm k}$ is an $M$-qubit state
and $\ket{\bm 0}$ is the state of $N$ auxiliary qubits.  We use the mapping
\bea
\ket{\bm k}\otimes \ket{\bm 0}&\to& \ket{\bm k} \otimes \ket{\bm 0} \hspace{1.35cm}{\rm if} \hspace{0.3cm} |{\bm k}|\not\in (k_-,k_+)\nonumber\\
\ket{\bm k}\otimes \ket{\bm 0}&\to& \ket{\bm k} \otimes \ket{|\bm k|-k_-} \hspace{0.3cm}{\rm if} \hspace{0.3cm} |{\bm k}|\in (k_-,k_+),
\eea
where $\ket{|\bm k|-k_-}=|0\rangle^{\otimes N-(|\bm k|-k_-)}\otimes|1\rangle^{\otimes |\bm k|-k_-}$ is a specific $N$-qubit state upon which $U(\vartheta)^{\otimes N}$ acts and $\ket{\bm k}$ is an $M$-qubit state upon which the identity acts (See Fig:~\ref{Fig_Spectrum}).  Notice that for $|\bm k|\in(k_-,k_+)$, the state $\ket{\bm k} \otimes \ket{|\bm k|-k_-}$
picks up the phase $e^{-i\vartheta(|\bm k|-k_-)}$, which is the correct phase up to an overall phase factor
$e^{i k_- \vartheta}$.  Moreover, the number of states with this phase factor corresponds to all $M$-bit strings
$\ket{\bm k}$ with Hamming weight $|\bm k|$, which is precisely the multiplicity of  $e^{-i |{\bm k}| \vartheta}$ for
$|\bm k|\in (k_-,k_+)$ in Eq.~\eqref{tildeV}.  All other states outside the bulk do not obtain a phase~\footnote{One may introduce an additional random phase $e^{-i\gamma |k|}$ with $\gamma\in(0,2\pi]$ whenever $|{\bm k}|\not\in(k_-,k_+)$, which guarantees that the protocol works equally well for all $\vartheta$.}.
For all other states $\{\ket{\bm k}\otimes \ket{\bm l}\}$ we can choose an arbitrary mapping to one of the other basis states such that the overall operator, $A$,
is unitary~\footnote{Note that it is sufficient for the map to perform the right action only when the $N$ auxiliary systems
are prepared in a given state.}. After application of $U^{\otimes N}$ to the last $N$ qubits, one only needs to undo the basis change by applying $A^\dagger$, see Fig. \ref{Fig_Setup}. The choice of $N = 2 \alpha M^\beta$ for $\beta >1/2$ ensures that the Jamio\l{}kowski fidelity is close to 1 in the limit of large $N,M$,
and hence super-replication with a rate of $O(N^2)$ is achieved. Note that one can indeed show that this rate is optimal. It is known that in state super-replication, the Heisenberg limit, i.e., a replication rate of $N^2$, is optimal~\cite{Ch:13}. As this also applies to the Choi-Jamio\l{}kowski state---which can be obtained deterministically from the unitary---any higher replication rate for unitaries would imply a corresponding higher rate for the state which is impossible~\footnote{G. Chiribella, private communication}.

Notice that in contrast to state super-replication our protocol works deterministically. This also holds when we apply the protocol to input states $|+\rangle^{\otimes M}$, which corresponds to the case of phase-covariant state cloning. The difference is that in our case the information on the unknown parameter, $\vartheta$, is encoded in the unitary operation and not in a particular state as is the case in~\cite{Ch:13}. Whereas standard cloning protocols deal with input states that are of tensor product structure, here it is possible to apply the unitary operations to general (entangled) states, which is effectively achieved by the mapping $A$. One can also directly adapt the protocol of~\cite{Ch:13} to accomplish deterministic state super-replication if we incorporate the filter into the state preparation procedure, prior to the application of the unitaries ---which imprint the state information--- and the cloning protocol.

We remark that our result can be generalized to arbitrary $d$-dimensional unitary operations
generated by a Hamiltonian with unknown interaction strength, $W(\vartheta)= \exp(-i\vartheta H)$ where
$H= \sum_j \lambda_j|\varphi_j\rangle\langle\varphi_j|$.
For $d > 2$, the relevant, likely subspace of $W(\vartheta)^{\otimes N}$ follows a multinomial, rather than a binomial,
distribution that converges to a multivariable Gaussian distribution centered at $p_k=\lambda_k N$. As long as the
Gaussian has a width of $O(\sqrt{N})$ in each dimension, the approximation is faithful. It follows that one can generate
an approximation of $W(\vartheta)^{\otimes N^2}$ from $W(\vartheta)^{\otimes N}$ in this case as well, where the required protocol is a
direct generalization of the one presented for $d=2$. The key ingredient is again the unitary operation $A$ where now
the tensor product of eigenstates $|\varphi_{\bm k}\rangle$, belonging to the likely subspace, are appropriately mapped
so that they pick up the correct phase factor when $W(\vartheta)^{\otimes N}$ is applied. As the spectral properties of
the Hamiltonian have no bearing in our argument, super-replication is possible for arbitrary Hamiltonians as well.

{\em Emulation of multi-qubit operations.---}
In a similar way one can also consider emulation of operations that depend on the same (unknown)
parameter, $\vartheta$, but act on different systems. For example, consider the operation
$V(\vartheta)=\exp(-i \vartheta H_V)$, where $H_V = \sum_{j=0}^{n-1} j |j\rangle\langle j|$
is the Hamiltonian acting on an $n$-level system, and the unitary operation $U(\vartheta)$ of Eq. \eqref{Utheta} acting
on a qubit. The above operations describe a spin-$(n-1)/2$ and a spin-$1/2$ particle coupled to the same magnetic field of
unknown strength $\vartheta$. Using the techniques established in the previous section it is straightforward to show
that a single use of $V(\vartheta)$ is sufficient to approximate $M$ uses of $U(\vartheta)$ whenever
$n = 2 \alpha M^\beta$ and $\alpha>0,\,\beta >1/2$. To see this first note that $V(\vartheta)$ and
$U(\vartheta)^{\otimes n}$ have the same spectrum; only the multiplicities of the various eigenvalues differ.
By attaching $M$ auxiliary qubits, on which the identity acts, one can construct a similar unitary operator to
$A$ above and obtain $U(\vartheta)^{\otimes n}$ \defn{exactly}~\footnote{We stress that there is no approximation
of $U(\vartheta)^{\otimes n}$ here as the entire spectrum of the latter can be obtained, not just the typical subspace.}.
Using the scheme described in the previous section we can now obtain an approximation of $U(\vartheta)^{\otimes n^2}$
from $n$ uses of $U(\vartheta)$.

The above result highlights an important equivalence between higher dimensional systems and the number of uses of a
unitary operator on a two-level system. One can trade a single use of a unitary acting on an $n$-level system for an
approximate  $n^2$ uses of a unitary operator acting on qubits.

So far we have considered that additional control is available only before and after the application of the unitary
operations. However, in many physically relevant situations, where $U(\vartheta)$ is generated by a Hamiltonian with
unknown interaction strength that is applied for a fixed time, additional control is available. In these cases one can
interject the Hamiltonian evolution with ultrafast control pulses thus modifying the effective evolution~\cite{Viola:98,*Viola:99}. This
technique, also known as ``bang-bang control", allows one to generate an effective Hamiltonian with a modified
spectrum. The use of bang-bang control techniques allows for more advanced emulation schemes. For example,
consider the $n$-fold degenerate Hamiltonian with eigenvalues $0, 1$.  Such a Hamiltonian describes, for example, the
spin and motional degrees of freedom of an electron, where the spin degrees of freedom are acted upon by the
Hamiltonian $H=\vartheta|1\rangle\langle 1|$---the same Hamiltonian that generates $U(\vartheta)$
in Eq.~\eqref{Utheta}---and $n$ motional degrees of freedom are acted on by the identity. Intermediate control pulses
allow one to modify the spectrum of the effective Hamiltonian such that it contains $n$ eigenstates, whose eigenvalues
are evenly gapped, and all but the ground state level are non-degenerate.  Up to a multiplicative factor of $n$, this is
the same spectrum as for the Hamiltonian $H_V$ above where the multiplicative factor leads to an evolution
$V(\vartheta/n)$ instead of $V(\vartheta)$.  Hence, one can use the same technique as before to obtain multiple
single-qubit operations. In fact, as $n$ can be freely chosen, we have that from a single application of $H$ for time
$t=1$, a single qubit operation $U(\vartheta)$, one can generate up to $n^2$ copies of an operation with reduced
strength $\vartheta/n$, i.e., $U(\vartheta) \to U(\vartheta/n)^{\otimes n^2}$.


{\em Links to quantum metrology.---}
We now discuss connections between the super-replication of unitary operations established above and
quantum metrology. The latter deals with optimally estimating an unknown parameter, $\vartheta$, by choosing an
optimal input state on which $\vartheta$ is imprinted, and reading out the desired information by means of an optimal
measurement~\cite{Giovannetti:04}. When the input state is a product state of $N$ qubits and the parameter, $\vartheta$, is imprinted
by applying the operation $U(\vartheta)$ of Eq.~\eqref{Utheta} on each qubit, then the achievable precision,
$\delta\vartheta$, in the estimation of $\vartheta$ is bounded by $\delta \vartheta \geq O(1/N)$, the
\defn{standard quantum limit}. When the $N$ qubits are prepared in an entangled state, however, an accuracy of
$\delta \vartheta = O(1/N^2)$ can be achieved.

Our super-replication procedure establishes an equivalence between different resources namely, $N$ uses of
$U(\vartheta)$ on an entangled input state of $N$ qubits, $N^2$ uses of $U(\vartheta)$ on the optimal product state of
$N^2$ qubits, and a single use of $V(\vartheta)=\exp(-i \vartheta H_V)$, where
$H_V = \sum_{j=0}^{N-1} j |j\rangle\langle j|$ acts on a single $N$-dimensional spin.

In particular, consider the case of quantum metrology where the input state comprises multiple qubits in the state
$|0_x\rangle=1/\sqrt{2}(|0\rangle + |1\rangle)$. This set-up corresponds precisely to phase covariant
cloning~\cite{Ch:13}, where the information on the phase is contained in the unitary $U(\vartheta)$, and,
in this case, the global fidelity of the cloned states is equivalent to the process fidelity. In \cite{Ch:13} it was shown that
the optimal super-replication strategy for states can produce at most $M=N^2$ copies, and saturates the Heisenberg
limit. Using our super-replication procedure for unitary operations, and applying it to the product input state
$|0_x\rangle^{\otimes N}$, one achieves the same, optimal, precision in quantum metrology as when
$U(\vartheta)^{\otimes N}$ act directly on the optimal entangled input state of $N$ qubits.

However, this does not guarantee that a high fidelity is achieved for all input states. The figure of merit for the super-replication procedure is the process fidelity, which provides a bound
on the average state fidelity averaged over all possible input states. We stress that the process fidelity is the standard
way of measuring the accuracy of operations and processes, and a high process fidelity implies a good approximation
of the process \cite{Ch:08,Gilchrist:05,*Dur:05}. On the one hand, there exist states where the fidelity exceeds the process fidelity, e.g. for any state of the form $|\psi\rangle=\sum_{|{\bm k}|\in(k_-,k_+)}\alpha_{\bm k}|{\bm k}\rangle$ the fidelity is one. On the other hand, there are also several input states for which the achievable fidelity is smaller than the process fidelity.

In fact, it turns out that the action of $U(\vartheta)^{\otimes M}$ is not appropriately mimicked for
input states that are themselves useful for parameter estimation, i.e., have a quantum Fisher information that is of
$O(M^2)$. This is to be expected, as otherwise the Heisenberg limit for metrology would be violated by combining
the super-replication of unitary operations as established here, and letting the protocol act on entangled input states.

Indeed, the protocol will not work for the optimal input state for quantum metrology,
$(|0\rangle^{\otimes M} + |1\rangle^{\otimes M})/\sqrt{2}$. Only random phases will be imprinted on both,
$|0\rangle^{\otimes M}$ and $|1\rangle^{\otimes M}$, rather than the required phases $0$ and $M\vartheta$.  By
construction, our super-replication protocol yields a faithful approximation only for the bulk of states where the Hamming
weight is approximately $M/2$, i.e., only for energy eigenstates with energy approximately $M/2 \pm \sqrt M$. All states
with a large support on this subspace have quantum Fisher information that scales only as $O(M)$. States with a
quantum Fisher information scaling as $O(M^2)$ are superpositions of eigenstates where the eigenvalues differ by
$O(M)$ \cite{Sk:14}. For all these states the proper phases are not reproduced by our super-replication protocol, however the relative volume of those states goes to zero with increasing $M$.

Finally, a single use of $V(\vartheta)$ on a $N$-dimensional spin also allows to mimic the action of $U(\vartheta)$ on $N^2$ product states $|0_x\rangle$, and hence to achieve the same precision in the estimation of $\vartheta$. One can trade between the number of levels and the number of copies of a two-level system.

{\em Conclusion and outlook.---}
We have demonstrated the deterministic super-replication of unknown unitary operations. For all operations generated
by a Hamiltonian with unknown interaction strength, one can produce up to $N^2$ copies of the operation using the
operation only $N$ times. This surprising result is in perfect agreement with similar effects in state super-replication and
quantum metrology. Whether a similar improvement can be obtained for arbitrary unitary operations of the group
$SU(2)$ remains an open question.

\textit{Note added.---} After completion of our work, the question regarding the super-replication of arbitrary unitary operations of the group SU(2) has been answered affirmatively in~\cite{Chiribella:14}

\textit{Acknowledgements.---} This work was supported by the Austrian Science Fund (FWF): P24273-N16 and the Swiss National Science Foundation grant P2GEP2\_151964.

\bibliographystyle{apsrev4-1}
\bibliography{superreplicate}

\end{document}